\documentclass[journal]{vgtc}                     

\onlineid{2105}

\vgtccategory{Research}

\title{GS-Surrogate: Deformable Gaussian Splatting for Parameter Space Exploration of Ensemble Simulations}

\author{
Ziwei Li$^{1}$,
Rumali Perera$^{1}$,
Angus G. Forbes$^{2}$,
Kenneth Moreland$^{3}$, 
David Pugmire$^{3}$, \\
Scott Klasky$^{3}$,
Wei-Lun Chao$^{1}$,
Han-Wei Shen$^{1}$ \\
\\
$^{1}$The Ohio State University  \quad
$^{2}$NVIDIA  \quad
$^{3}$Oak Ridge National Laboratory  
}

\authorfooter{
  \item
  	Ziwei Li, Rumali Perera, Wei-Lun Chao, and Han-Wei Shen with The Ohio State University.
  	E-mail: {\{li.5326, perera.62, chao.209, shen.94\}@osu.edu}
  \item 
    Angus G. Forbes with NVIDIA. E-mail: aforbes@nvidia.com
  \item
    Kenneth Moreland, David Pugmire, and Scott Klasky with Oak Ridge National Laboratory. 
    E-mail: {\{morelandkd, dpn, klasky\}@ornl.gov}
}

\abstract{%
Exploring ensemble simulations is increasingly important across many scientific domains. However, supporting flexible post-hoc exploration remains challenging due to the trade-off between storing the expensive raw data and flexibly adjusting visualization settings. Existing visualization surrogate models have improved this workflow, but they either operate in image space without an explicit 3D representation or rely on neural radiance fields that are computationally expensive for interactive exploration and encode all parameter-driven variations within a single implicit field.
In this work, we introduce \ours, a deformable Gaussian Splatting-based visualization surrogate for parameter-space exploration. Our method first constructs a canonical Gaussian field as a base 3D representation and adapts it through sequential parameter-conditioned deformations. By separating simulation-related variations from visualization-specific changes, this explicit formulation enables efficient and controllable adaptation to different visualization tasks, such as isosurface extraction and transfer function editing.
We evaluate our framework on a range of simulation datasets, demonstrating that \ours enables real-time and flexible exploration across both simulation and visualization parameter spaces. 
}


\graphicspath{{figs/}{figures/}{pictures/}{images/}{./}} 

\usepackage{tabu}                      
\usepackage{booktabs}                  
\usepackage{lipsum}                    
\usepackage{mwe}                       
\usepackage{ccicons}                   

\usepackage{mathptmx}                  
\usepackage{makecell}
\usepackage{multirow}
\usepackage{array} 
\usepackage{amsmath}
\usepackage{soul}

\usepackage{xcolor}

\usepackage{amsmath,amsfonts,bm}

\def\eqref#1{equation~\ref{#1}}

\def\1{\bm{1}}

\DeclareMathAlphabet{\mathsfit}{\encodingdefault}{\sfdefault}{m}{sl}
\SetMathAlphabet{\mathsfit}{bold}{\encodingdefault}{\sfdefault}{bx}{n}

\usepackage{amssymb}
\usepackage{amsmath,amsfonts}
\usepackage{amsopn}
\usepackage{bm} 
\usepackage{multirow}
\usepackage{flushend}
\usepackage{tabularx}
\usepackage{xspace}

\def\eqref#1{equation~\ref{#1}}

\def\1{\bm{1}}

\DeclareMathAlphabet{\mathsfit}{\encodingdefault}{\sfdefault}{m}{sl}
\SetMathAlphabet{\mathsfit}{bold}{\encodingdefault}{\sfdefault}{bx}{n}
\newcommand{\eat}[1]{}

\newcommand{\ours}{{GS-Surrogate}\xspace}

\newcommand{\eg}{{\em e.g.}}
\newcommand{\ie}{{\em i.e.}}

\begin{document}



\maketitle

\section{Introduction}
\label{intro_sec}

Ensemble simulations are essential in many scientific domains to investigate how physical systems evolve under varying conditions. To gain deeper scientific insights, scientists usually run simulations across a broad range of parameter settings. However, this exploration process often faces two long-standing challenges. 
First, saving the high-resolution simulation outputs leads to massive I/O operations and substantial storage overhead. 
Second, although in-situ rendering~\cite{ma2009situ,bauer2016situ} can reduce storage overhead, it typically fixes the rendering configurations and limits flexibility for the post-hoc analysis. Ensemble simulation exploration requires interactively adjusting both simulation parameters and visualization settings (\eg, viewpoints and transfer functions), but existing pipelines often enforce a trade-off between storage efficiency and support for post-hoc analysis.
\par
To address these challenges, visualization surrogate models~\cite{he2019insitunet, han2022coordnet, yao2025visnerf} have been proposed to synthesize images directly from simulation and visualization parameters. Despite their storage efficiency, most existing methods rely on learning end-to-end mappings from input parameters to 2D images. Without explicitly modeling the underlying 3D structures, they are less effective at learning geometrically coherent and view-consistent representations for volumetric data.
\par
Recently, neural radiance fields (NeRF)~\cite{mildenhall2021nerf,chen2022tensorf} have been incorporated to capture geometry-aware scenes for scientific visualization. However, NeRF-based visualization surrogates~\cite{yao2025visnerf} often suffer from significant rendering cost due to the dense ray sampling and repeated queries along each ray. In this case, interactive exploration becomes even more challenging for ensemble simulations. For example, uncovering a complex scientific phenomenon often requires a sequence of interactive operations, such as navigating the viewpoint, sweeping through the simulation conditions, and then adjusting the transfer functions. Under such workflows, it is difficult to deploy neural rendering-based surrogates for real-time parameter-space exploration.
Moreover, these methods usually encode all parameter-dependent variations into a single unified implicit field. Such a formulation can introduce several challenges for reliable post-hoc analysis. 
First, since structural variations across different parameter spaces are entangled within the same representation, learning one type of variation may interfere with another. For example, fitting the appearance changes caused by different transfer functions could affect the geometric differences learned across different ensemble members, which may lead to rendering artifacts.
Second, scientific data often contains sparse and localized features, whereas grid-based radiance fields allocate capacity uniformly across the entire volume, limiting their ability to capture fine-grained structures with high fidelity. 
Third, when new visualization settings are introduced during the post-hoc exploration, these methods typically require retraining the entire representation from scratch.
\par
An ideal visualization surrogate should produce view-consistent renderings from arbitrary viewpoints, support real-time parameter-space exploration, and remain flexible enough to adapt to different visualization tasks.
3D Gaussian Splatting (3DGS)~\cite{kerbl20233d} has recently emerged as a promising approach for achieving this goal.
By learning a set of Gaussian primitives, 3DGS provides a geometry-aware and highly adaptive representation, which is particularly appealing for scientific visualization.
In addition, 3DGS offers a more efficient rendering pipeline than NeRF-based methods, making it well-suited for supporting real-time exploration. However, effectively extending 3DGS for parameter-space exploration is non-trivial for two main reasons.
First, the underlying Gaussian primitives need to provide sufficient spatial coverage to capture both shared and member-specific features across the ensemble. 
Second, existing dynamic 3DGS approaches~\cite{wu20244d, yang2024deformable} are primarily designed for learning smooth temporal variations. In contrast, scientific volumes may present significant changes in terms of both structure and appearance under different parameter settings. 
\par
In this work, we introduce \textbf{\ours}, a deformable Gaussian Splatting-based visualization surrogate, built upon a reusable canonical Gaussian field for post-hoc exploration of ensemble simulations. 
Specifically, \ours first constructs a canonical Gaussian field to capture the essential geometric structures required for modeling all ensemble members. 
We then propose a two-level deformation framework that first adapts the canonical representation to simulation-parameter variations and then to different visualization tasks. We show that the proposed modular design enables flexible, high-quality, and real-time exploration across both simulation and visualization parameter spaces.
Furthermore, in each deformation model, we explicitly decouple geometry-related variations from appearance-related adaptations, enabling more effective and controllable exploration across different tasks.
\par
In summary, the main contributions of this work are as follows:
\begin{itemize}
    [nolistsep]
    \item Presenting \ours, a novel surrogate model that leverages deformable 3D Gaussian representation to support view-consistent, geometry-aware rendering and interactive post-hoc exploration for ensemble simulations.
    \item Introducing a reusable canonical Gaussian representation with parameter-conditioned deformation, where a modular design explicitly separates simulation-conditioned adaptation from visualization-related mappings. 
    \item Supporting high-quality, real-time exploration of ensemble simulations across the parameter space.
\end{itemize}

\section{Related Work}
\label{sec:related_work}

In this section, we review prior work on image-centric in situ visualizations, surrogate models for parameter-space exploration, and existing methods on dynamic scene rendering.

\textbf{Image-Centric In Situ Visualization.}
In situ approaches that store images rather than raw simulation data have been a promising direction for managing the I/O and storage demands of large-scale scientific simulations. Ma \cite{ma2009situ} identified the challenges for in situ visualization at extreme scales, noting that data reduction is essential when I/O bandwidth cannot keep pace with simulation speeds. Ahrens et al. \cite{ahrens2014image} developed the Cinema framework, which records parameterized image databases in situ, allowing interactive post-hoc analysis. Biedert and Garth \cite{biedert2015contour} combined topological analysis with image-based data representations to allow post-hoc exploration flexibility while avoiding the need to store the full volumetric data. Frey et al. \cite{frey2013explorable} proposed Volumetric Depth Images, a compact image-based representation that captures depth and color information during raycasting and can subsequently be rendered from arbitrary viewpoints, allowing flexible post-hoc view exploration. However, these approaches are fundamentally constrained to the parameter configurations sampled at simulation time, and cannot generalize to unseen simulation conditions without re-running the underlying simulation.

\textbf{Surrogate Models for Parameter Space Exploration.}
He et al. \cite{he2019insitunet}  introduced InSituNet, a convolutional regression model trained on in situ collected image databases to predict visualization results from joint simulation and visualization parameters. Berger et al. \cite{berger2018generative} proposed a deep learning approach for transfer function design of volume renderings using generative adversarial networks. Recently, Yao et al. \cite{yao2025visnerf} introduced ViSNeRF, which constructs a multidimensional neural radiance field from sparse in situ collected images to support viewpoint synthesis across transfer functions, isovalues, and simulation parameters. Although these image-based surrogates avoid storing raw simulation data, they operate in 2D image space or rely on slow implicit rendering, limiting either their generalization capacity or their suitability for real-time exploration. In contrast, our work addresses both limitations by adopting an explicit Gaussian primitive representation that is trained purely from images while enabling real-time rendering.
\par
In contrast to image-based surrogates, which operate in the visual domain, data-based approaches act directly on the simulation's physical variables. Shi et al. \cite{shi2022vdl} introduced VDL-Surrogate, which encodes raw volumetric data into view-dependent latent representations and decodes them into high-resolution images conditioned on simulation and viewpoint parameters. Shi et al. \cite{shi2022gnn} proposed GNN-Surrogate, a hierarchical graph neural network that predicts simulation output fields on unstructured meshes, allowing scientists to apply arbitrary transfer functions post-hoc. However, these approaches require access to and storage of raw volumetric simulation data throughout training, reintroducing significant I/O and storage overhead.

Existing approaches for parameter-space exploration of ensemble simulations fall into two main categories. The first one relies on standard high-dimensional data visualization techniques applied directly to collected ensemble inputs and outputs. Parallel coordinates \cite{obermaier2015visual, wang2016multi}, scatter plots \cite{matkovic2009interactive, orban2018drag}, radial plots \cite{bruckner2010result, chen2015uncertainty}, glyphs \cite{bock2015visual}, and matrix-based views \cite{poco2014visual} have all been used to analyze relationships across ensemble members. A fundamental limitation shared by all these methods is that analysis remains confined to parameter configurations that were explicitly simulated.
The second category, including our \ours, uses surrogate models to predict outcomes at new, unsampled parameter configurations, extending exploration beyond the limits of the collected ensemble. Rather than being confined to pre-simulated settings, scientists can freely navigate the parameter space to investigate how changes in physical conditions affect the simulation output, supporting tasks such as sensitivity analysis and feature tracking across the parameter space.

\textbf{Dynamic Scene Rendering.}
Recent work on dynamic scene representations can be generally divided into NeRF-based and Gaussian splatting-based methods. For example, K-Planes \cite{fridovich2023k} and HexPlane \cite{cao2023hexplane} factorize the 4D spacetime volume into compact planar representations to allow efficient dynamic novel view synthesis. More recently, dynamic 3D Gaussian Splatting \cite{wu20244d, yang2024deformable, yang2023real, bae2024per,li2024st} extends the explicit Gaussian primitive representation to dynamic scenes by learning per-Gaussian deformation fields that warp a canonical set of Gaussians across time, allowing real-time rendering of dynamic sequences. Along this direction, \cite{lu20243d} incorporates 3D geometry awareness into the deformation framework to improve dynamic view synthesis. However, all these methods treat time as the axis of variation and are designed for monocular video reconstruction. In contrast, our work extends the Gaussian splatting framework to ensemble simulation analysis.

\section{Background: 3D Gaussian Splatting}
\label{sec:background}

In this section, we review the basic concepts of 3D Gaussian Splatting (3DGS)~\cite{kerbl20233d}, including the representation of 3D Gaussian primitives and the differentiable splatting-based rendering process.
\par
3DGS provides an explicit representation of a 3D scene using a set of anisotropic Gaussian primitives. Specifically, each 3D Gaussian is defined by a mean position $\mu \in \mathbb{R}^{3}$ and a 3D covariance matrix $\Sigma \in \mathbb{R}^{3\times3}$:
\begin{equation}
    G(\mathbf{x}) = \exp\left(
    -\frac{1}{2}
    (\mathbf{x}-{\mu})^{\top}
    \Sigma^{-1}
    (\mathbf{x}-{\mu})
    \right),
\label{eq:gs_primitives}
\end{equation}
where $\mathbf{x} \in \mathbb{R}^{3}$ denotes a 3D coordinate. To facilitate the optimization and ensure that the covariance matrices are positive semi-definite, 3DGS reparameterizes $\Sigma$ using a 3D vector $s \in \mathbb{R}^{3}$ for scaling and a quaternion $q \in \mathbb{R}^{4}$ for rotation. The covariance matrix of an anisotropic Gaussian is then expressed as $\Sigma = R S S^{T} R^{T}$, where $R$ denotes the rotation matrix converted from $q$ and $S$ is the diagonal scaling matrix defined by the 3D scaling vector $s$.
\par
Moreover, to model the view-dependent appearance, each Gaussian also stores a set of spherical harmonic (SH) coefficients~\cite{kerbl20233d,fridovich2022plenoxels} and an opacity value $o \in \mathbb{R}$. Therefore, every 3D Gaussian primitive is parameterized by five attributes: position $\mu_i$, scaling $s_i$, rotation $q_i$, appearance coefficients $c_i$, and opacity $o_i$.
\par
In contrast to Neural Radiance Fields (\ie, NeRF~\cite{mildenhall2021nerf}), which rely on a continuous dense representation and perform volume rendering via ray marching, 3DGS is a point-based modeling approach that enables Gaussian primitives to be efficiently rasterized for real-time rendering. 
\par
In general, the rendering process of 3DGS contains two key steps: \textit{splatting} and \textit{$\alpha$-blending}~\cite{kerbl20233d}. In the splatting stage, each 3D Gaussian is projected onto the 2D image space. With $W$ as the viewing transform and $J$ as the Jacobian from the affine linearization of the projective transformation, the camera space covariance matrix $\Sigma'$ is approximated as $\Sigma' = J W \Sigma W^{\top} J^{\top}$. In the $\alpha$-blending stage, for each image pixel, all the projected 2D Gaussians overlapping with that pixel are first sorted by depth. The final color $C$ of that pixel, denoted by $\mathbf{p}$, is then computed by compositing all these overlapping Gaussians as: 
\begin{equation}
    C(\mathbf{p}) = \sum_{i \in N} c_i \alpha_i \prod_{j=1}^{i-1} (1 - \alpha_j), 
\end{equation}
where $\alpha_i$ and $c_i$ denote the opacity and color of the $i$-th Gaussian contributed at that pixel. The term $\prod_{j=1}^{i-1} (1 - \alpha_j)$ represents accumulated transmittance based on the opacity of all the Gaussians in front of the $i$-th Gaussian along the viewing direction.

\section{Framework Overview}
\label{sec:dataset_generation}

\begin{figure}[htbp]
    \centering
    \noindent\includegraphics[width=0.99\linewidth]{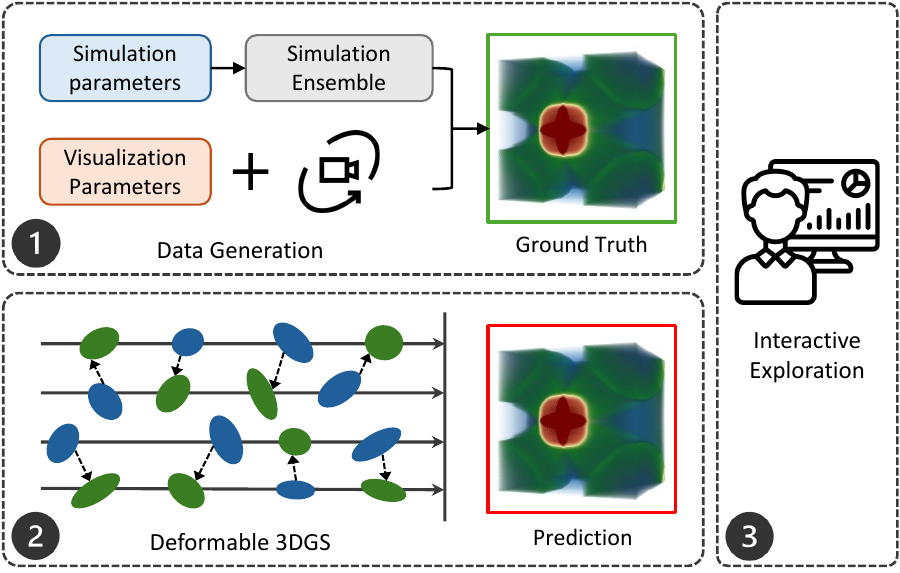}
    \caption{Overall pipeline of \ours. (1) Training images are generated through in situ visualization. (2) A deformable 3DGS-based visualization surrogate is trained offline using the image collection. (3) At inference time, scientists can interactively explore the rendering results across the parameter space.}
    \label{fig:frame_overview}
\end{figure}

\noindent \autoref{fig:frame_overview} illustrates the overall pipeline of our framework, which consists of three main components: (1) training data generation, (2) learning a deformable 3D Gaussian splatting-based surrogate model, and (3) interactive exploration of the ensemble simulations.

\textbf{Dataset Generation.}
To create the multi-view training images, we systematically vary three types of parameters: simulation parameters \textbf{$\text{P}_\text{sim}$}, visualization parameters \textbf{$\text{P}_\text{vis}$}, and view parameters. 
\textit{Simulation parameters} are represented as multivariate vectors that encode the physical conditions under which each simulation instance is generated, with bounds predefined by the domain experts.
\textit{Visualization parameters} describe the rendering operations applied to each simulation output, such as isosurface extraction at specific isovalues or volume rendering under varying transfer functions. 
For \textit{viewpoint selection}, we adopt an icosphere-based sampling strategy \cite{yao2025visnerf, shi2022gnn} that provides uniform coverage around the volume data. Camera positions are held constant across all ensemble members for both volume rendering and isosurface rendering tasks.
By combining these parameters, we obtain a collection of parameter-image pairs as the ground truth used for training our visualization surrogate. Further details on the datasets are provided in \cref{sec:results}. 
\par
\textbf{Deformable GS-based Surrogate.} 
The key component of our framework is a deformable 3DGS model that learns to synthesize visualization results conditioned on simulation parameters and visualization settings. We first construct a canonical set of Gaussian primitives \textit{G} that serves as the base representation for all ensemble members. Then, a deformation process is learned to transform these Gaussian primitives into a member-specific representation: \textit{G'}$ = \mathcal{F}(\textit{G}, {\textit{P}_\text{sim}}, \textit{P}_\text{vis})$. The deformed Gaussians \textit{G'} are then rendered through differentiable rasterization. The entire pipeline is trained by minimizing the reconstruction loss between the rendered images and the ground truth. At inference time, given an arbitrary combination of simulation parameters and visualization configurations, the deformation network produces the corresponding Gaussians, which can be rendered from any viewpoints in real time.
\par
\textbf{Interactive Exploration.}
Once trained, our \ours enables scientists to interactively explore the ensemble simulation by freely adjusting the viewpoints, simulation parameters, and visualization settings, without re-running the underlying simulations.
\par
The architecture and training procedure of our deformable GS-based surrogate are introduced in \cref{sec:method}. We further present case studies in \cref{sec:exploration} to demonstrate how \ours can support interactive exploration of ensemble simulations.

\section{\ours}
\label{sec:method}

\begin{figure}[htbp]
    \centering
    \noindent\includegraphics[width=1.0\linewidth]{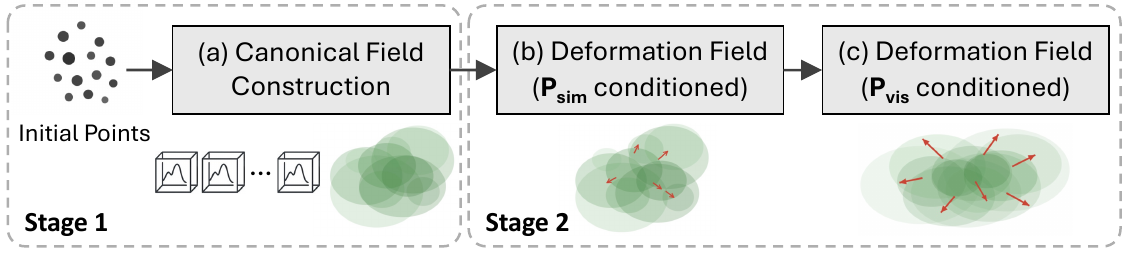}
    \caption{Overview of the two-stage training pipeline of \ours. The first stage optimizes a set of canonical Gaussian primitives. In the second stage, the deformation fields conditioned on simulation and visualization parameters are trained sequentially.}
    \label{fig:two_stages}
\end{figure}

\autoref{fig:two_stages} illustrates the overall pipeline of \ours. During training, our framework consists of two major stages: (1) constructing a set of canonical 3D Gaussian primitives, and (2) learning the parameter-conditioned deformation field that transforms these canonical Gaussians to match specific simulation and visualization configurations.
During inference, given any combination of parameters, \ours deforms the canonical Gaussians accordingly and renders the scene in real time using the differentiable splatting process described in \cref{sec:background}.
\par
Our framework is designed with two considerations. First, by decomposing the entire learning process into canonical Gaussian reconstruction and parameter-conditioned deformation, the model can \textit{reuse} a set of shared geometric representations \textit{across different simulation parameters}. In contrast to neural radiance fields, which implicitly encode both geometry and appearance into a single implicit representation, our explicit Gaussian primitives allow the underlying geometric structures to be shared and efficiently adapted through deformation. 
Second, this explicit representation naturally supports \textit{task decomposition}. Since the canonical field is learned independently of visualization parameters, the same underlying geometric representations can be efficiently adapted to \textit{different visualization tasks}, such as transfer function editing or isosurface extraction.

\subsection{Stage 1: Canonical Field Construction}
\label{sec:canonical}

The goal of the canonical field is to construct a set of 3D Gaussian primitives that can be shared across all the ensemble members, such that the deformation network can model each target field by solely adjusting these Gaussians. To serve as an effective base representation, this canonical field is expected to provide sufficient spatial coverage. Moreover, it should capture not only the geometric structures shared by the ensemble, but also enough local features to facilitate the reconstruction of member-specific structures in the second stage. 
\par
To initialize the canonical field, we first apply Structure-from-Motion (SfM) to a representative member selected from the training ensemble, \eg, the one closest to the mean in the simulation parameter space. This step produces a sparse point cloud that serves as the initial positions and colors of the Gaussian primitives. We then optimize the canonical field using images randomly sampled across training ensemble members and viewpoints for a fixed number of iterations.
A key component in constructing this canonical field is the densification strategy. Specifically, at regular iterations, Gaussians with opacity below a predefined threshold are pruned, while those in under-reconstructed regions are densified by splitting or cloning the existing Gaussians based on their positional gradients. 
This iterative process is particularly important in our setting, as randomly sampled views from different ensemble members prompt the model to create additional Gaussians in those under-represented regions. To better accommodate a wide range of simulation parameters, we employ a slightly lower gradient threshold to encourage a denser representation.


\subsection{Stage 2: Parameter-Conditioned Deformation}
\label{sec:deformation}
\subsubsection{Model Overview}
\label{sec:model_overview}
After the canonical field has been constructed, the goal of the second stage is to learn a deformation field that transforms the Gaussian primitives to match specific simulation and visualization parameters. As illustrated in \autoref{fig:two_stages}-(b) and (c), we decompose the deformation learning into two sequential steps.
\par
First, we train a network $F_{\text{sim}}$ that adapts the canonical Gaussians to each simulation condition. Specifically, $F_{\text{sim}}$ takes the position of each canonical Gaussian $\mu$ together with the simulation parameter $\text{P}_{\text{sim}}$ as input, and predicts offsets for the Gaussian attributes. Formally, the simulation parameter-conditioned deformation of Gaussian primitives is defined as:
\begin{equation}
     {F_\text{sim}}({\mu}, \text{P}_\text{sim}) = (\Delta{\mu}^\text{p}, \Delta{s}^\text{p}, \Delta{q}^\text{p}, \Delta{c}^\text{p}, \Delta{o}^\text{p}),
     \label{eq:deformation_sim}
\end{equation}
where the predicted offsets correspond to the Gaussian position, scaling, rotation, color, and opacity attributes, respectively.
\par
These offsets capture three types of physical variations in the scientific fields. First, the position offset $\Delta{\mu}$, rotation offset $\Delta{q}$, and scaling offset $\Delta{s}$ 
For example, for global ocean simulations, increasing the wind stress parameter can change the ocean current structure and even expand the spatial extent of a localized temperature field. 
Second, the opacity offset handles the \textit{visibility changes}. By controlling the visibility of Gaussians, the deformation network can model the topological changes in the field. For instance, in cosmological simulations, a filament structure may disappear under certain simulation configurations.  
Third, the color offset $\Delta{c}$ captures the \textit{appearance variations}. Even though the geometric structure may remain the same at a fixed location, its visual appearance can vary due to changes in the underlying scalar field.
\par
In the second step, the goal is to further deform the same set of Gaussians to support a specific visualization task, such as isosurface extraction. The architecture of $F_\text{vis}$ is similar to $F_\text{sim}$, except that it takes the already deformed Gaussian position and an additional visualization parameter $P_\text{vis}$ as input. The second deformation process is formulated as:
\begin{equation}
     {F_\text{vis}}({\mu'}, \text{P}_\text{sim}, \text{P}_\text{vis}) = (\Delta{\mu}^\text{v}, \Delta{s}^\text{v}, \Delta{q}^\text{v}, \Delta{c}^\text{v}, \Delta{o}^\text{v}),
     \label{eq:deformation_vis}
\end{equation}
where $\mu'=\mu+\Delta{\mu}^{p}$. Given a parameter setting, the final deformed Gaussians are obtained by sequentially applying the predicted offsets to the canonical field. The resulting Gaussians are then rendered via differentiable splatting to produce the final visualization.
\par
The architecture of each deformation model consists of two main components: (1) the \textit{spatial-parameter encoders} that jointly learn a shared parameter-conditioned feature representation, and (2) the \textit{multi-head feature decoders} that predict the attribute offsets. 
Although this sequential design may appear to introduce additional computational overhead during both training and inference, our modular formulation keeps the visualization deformation model lightweight. In particular, $F_{\text{vis}}$ uses a lighter feature decoder and can optionally disable certain decoder heads depending on the target visualization task. The details of each component in $F_{\text{sim}}$ and $F_{\text{vis}}$ are described in the following subsections.

\subsubsection{Spatial-Parameter Encoder}
\label{sec:encoder}

The encoder module aims to map the Gaussian positions and the parameter configurations into a joint feature representation that captures how the spatial feature of each Gaussian varies under specific parameter settings. 
\par
\begin{figure}[htbp]
    \centering
    \noindent\includegraphics[width=0.99\linewidth]{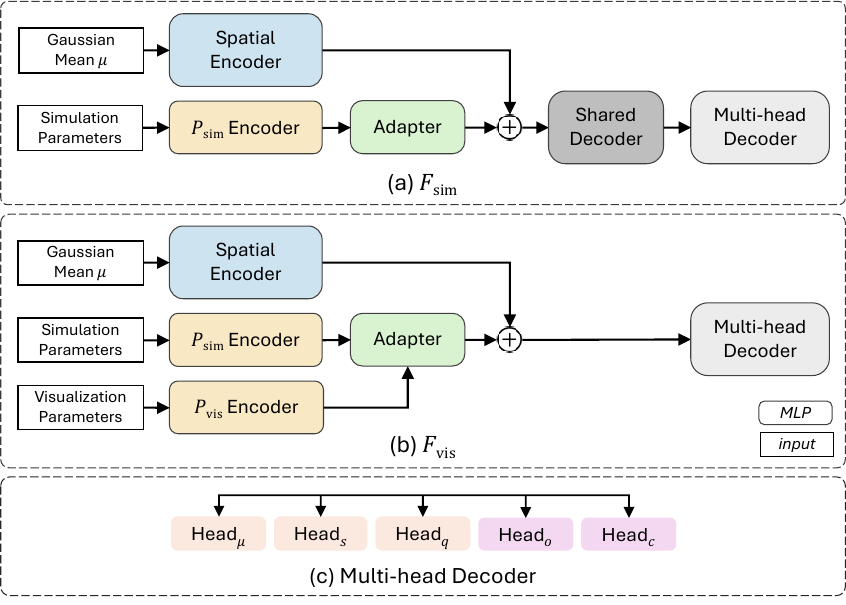}
    \caption{Architecture of the two deformation networks, $F_{\text{sim}}$ and $F_{\text{vis}}$, trained in the second stage.}
    \label{fig:deformation_network}
\end{figure}
As illustrated in \autoref{fig:deformation_network}, our encoder module contains three types of encoder branches. 
The first branch is the \textit{spatial encoder} $f_{\theta_\text{sp}}$, which embeds the canonical Gaussian position $\mu$ using positional encoding followed by a small MLP to produce the spatial feature $\mathbf{z}_\text{x}$, where $\mathbf{z}_\text{x} = f_{\theta_\text{sp}}(\gamma({{\mu}}))$. 
The second branch is the \textit{simulation-parameter encoder} $f_{\theta_\text{sim}}$, which embeds the multivariate $P_\text{sim}$ as conditioning features, \ie, $\mathbf{z}_{\text{p}_\text{sim}} = f_{\theta_\text{sim}}(\gamma({\text{P}_\text{sim}}))$. 
The third branch is the \textit{visualization-parameter encoder} $f_{\theta_\text{vis}}$, which exists only in $F_{\text{vis}}$ shown in \autoref{fig:deformation_network}-(b). 
In our implementation, for the isosurface extraction task, we encode a single isovalue. 
For transfer function (TF) editing, we focus on the opacity mapping defined by a set of control points in the value-opacity space. The TF is discretized into 256 control points, while the editing operation is parameterized by the coordinates of two movable control points. Each new TF instance is then represented by the signed displacement of these control points relative to a predefined base TF.
\par
To fuse the spatial features with the conditioning information,  we introduce a lightweight \textit{adapter} network ${f}_{\theta_\mathbf{A}}$. Specifically, the adapter MLP takes the parameter embedding as input and predicts a residual feature vector $\Delta{\mathbf{z}_\text{x}} = {f}_{\theta_\mathbf{A}}(\mathbf{z}_{\text{p}_\text{sim}})$, which is added to the initial spatial feature: $\mathbf{z'}_\text{x} = \mathbf{z}_\text{x} + \Delta{\mathbf{z}_\text{x}}$. The resulting feature $\mathbf{z'}_\text{x}$ is then passed to the decoder module. This residual design not only stabilizes the training process but also preserves the original spatial information by introducing only a small feature perturbation through parameter conditioning.

\subsubsection{Multi-head Decoder}
\label{sec:decoder}

The decoder module aims to predict two types of Gaussian attribute offsets: geometric offsets and appearance offsets. 
For the simulation parameter-conditioned deformation model $F_{\text{sim}}$ (see \autoref{fig:deformation_network}-(a)), we first employ a \textit{shared decoder backbone $f_{\theta_\mathbf{D}}$} to learn a common deformation pattern given the feature vector $\mathbf{z'}_\text{x}$. This shared feature is then passed to multiple lightweight prediction \textit{heads} $\Phi$ (see \autoref{fig:deformation_network}-(c)), which estimate the offsets of individual Gaussian attributes.
\par
Formally, the deformation of each attribute is computed as follows: position $\Delta{\mathbf{\mu}}=\Phi^{\mu}({f_{\theta_\mathbf{D}}(\mathbf{z'}_\text{x})})$, scaling $\Delta{\mathbf{s}}=\Phi^{s}({f_{\theta_\mathbf{D}}(\mathbf{z'}_\text{x})})$, rotation $\Delta{\mathbf{q}}=\Phi^{q}({f_{\theta_\mathbf{D}}(\mathbf{z'}_\text{x})})$, view-dependent color $\Delta{\mathbf{c}}=\Phi^{c}({f_{\theta_\mathbf{D}}(\mathbf{z'}_\text{x})})$, and opacity $\Delta{\mathbf{o}}=\Phi^{o}({f_{\theta_\mathbf{D}}(\mathbf{z'}_\text{x})})$. 
Each prediction head is implemented as a lightweight MLP.
\par
For the deformation model $F_{\text{vis}}$ (see \autoref{fig:deformation_network}-(b)), we remove the shared decoder backbone to obtain a lighter architecture. Moreover, for certain visualization tasks, such as TF editing, since Gaussian geometry has already been well adapted in the first deformation step, $F_{\text{vis}}$ only needs to model the changes in color and opacity. In this case, the three geometry-related heads are disabled, and only the appearance-related heads are retained to predict opacity and color offsets. This modular design not only improves computational efficiency but also helps reduce overfitting by avoiding unnecessary deformation of Gaussian attributes.

\subsection{Optimization}

\subsubsection{Loss Functions}
Both training stages are supervised by comparing the rendered images with the ground-truth views using an $L_1$ reconstruction loss combined with a structural similarity loss that penalizes perceptual differences: 
\begin{equation}
    L_{\text{color}} = (1-\lambda)L_1(I,\hat{I}) + \lambda (1-\text{SSIM}(I,\hat{I})),
\end{equation}
where $I$ and $\hat{I}$ denote the rendered image and ground-truth image respectively. Following the setting in the original 3DGS~\cite{kerbl20233d}, we set $\lambda = 0.2$ in all experiments.
\par
Furthermore, to encourage smooth deformations and ensure a stable training process in the second stage, we introduce a regularization term that constrains the magnitude of the predicted offsets and prevents the network from producing excessively large deformations. Given the offset predictions defined in \cref{sec:model_overview}, the deformation regularization term is formulated as:
\begin{equation}
    L_{\text{deform}} = \|\Delta \mu\|_2^2 + \|\Delta s\|_2^2 + \|\Delta q\|_2^2,
\end{equation}
where $\Delta o$ and $\Delta c$ can also be regularized when the appearance-related deformation is enabled.
The final training objective is defined as:
\begin{equation}
    L = L_{\text{color}} + \lambda_{\text{deform}} L_{\text{deform}},
\end{equation}
where $\lambda_{\text{deform}}$ control the strength of the deformation regularization.


\subsubsection{Training Strategy}

To ensure that the canonical Gaussians can be effectively adapted to different visualization tasks in the second stage, we introduce several training strategies.
\par
\paragraph{Canonical field fine-tuning.} When training the first deformation model ${F}_\text{sim}$, we allow the canonical Gaussians to be jointly updated with the deformation network, but with a significantly reduced learning rate ($0.001\times$ the deformation learning rate). This design preserves the geometric foundation built in the first stage while still providing sufficient flexibility for the model to accommodate different simulation conditions.
\par
\paragraph{Selective freezing in the second deformation step.} After ${F}_\text{sim}$ has been learned, the canonical field is completely frozen during the training of the second deformation model ${F}_\text{vis}$. Furthermore, depending on the target visualization task, we choose different fine-tuning strategies for ${F}_\text{sim}$. Specifically, for tasks requiring substantial geometric changes, such as isosurface extraction under different isovalues, we allow ${F}_\text{sim}$ to be further fine-tuned with ${F}_\text{vis}$. In contrast, for tasks like transfer function editing, where the underlying geometry remains unchanged, we freeze ${F}_\text{sim}$ and train only the second deformation model. In this case, ${F}_\text{vis}$ serves as a lightweight appearance adapter that only learns the changes in color and opacity introduced by the new TFs. This training strategy leads to faster convergence and more stable training.
\par
\paragraph{Hard example sampling.} During the optimization of both deformation models, we observe that images under certain viewpoints and parameter settings contain more complex structures that are difficult for the model to learn. To improve performance on these challenging examples, we employ a weighted sampling strategy where images with larger reconstruction errors are sampled more frequently. This strategy encourages the model to focus on under-fitted views and facilitates a balanced reconstruction quality across different conditions.

\section{Results}
\label{sec:results}

\subsection{Datasets}

We evaluate our method on four ensemble simulation datasets covering both volume rendering and isosurface rendering tasks.
Across all datasets, camera positions are derived from the vertices of a subdivided icosahedron at refinement level 5, resulting in 252 viewpoints that provide balanced coverage around the volume data. These viewpoints are split evenly, with 126 used for training and 126 reserved for testing. 
Additional dataset statistics are summarized in \autoref{tab:datasets}.

\textbf{Nyx} \cite{almgren2013nyx} is a cosmological hydrodynamics simulation developed at Lawrence Berkeley National Laboratory. We examine three physical parameters that govern the large-scale structure of the universe: total matter density $\mathit{OmM} \in [0.12, 0.155]$, baryon density $\mathit{OmB} \in [0.0215, 0.0235]$, and the Hubble constant $h \in [0.55, 0.85]$. The simulation produces a $512 \times 512 \times 512$ scalar volume representing the logarithmic dark matter density field. 
\par
To support the transfer function editing experiments, each ensemble member is visualized in situ via volume rendering under 77 distinct transfer functions. These transfer functions are defined by two fixed control points at $(s=0,\, o=0)$ and $(s=1,\, o=1)$, where $s$ denotes the scalar value and $o$ denotes the opacity, and two movable interior control points $c1$ and $c2$. The 77 training TFs are constructed as follows: a base TF is defined by the default positions of $c1$ and $c2$; 64 TFs are generated by combinatorially displacing both control points across all combinations of their scalar and opacity steps and 12 additional TFs are generated by varying only either the scalar or the opacity of one control point at a time capturing finer marginal variations in the TF space. It is important to note that, for each TF, only a randomly selected 25\% subset of the ensemble members is used during training. This greatly improves training efficiency by avoiding the need to render all members under every transfer function. Even with this reduced subset, \ours is able to learn a representative deformation model that generalizes smoothly to unseen transfer functions at inference time. The resulting images have a resolution of $256 \times 256$. We use 100 parameter configurations for training and 30 for testing.

\textbf{MPAS-Ocean} \cite{ringler2013multi} is a global ocean circulation model developed at Los Alamos National Laboratory. We investigate four parameters suggested by domain scientists that influence large-scale ocean dynamics: Bulk Wind Stress Amplification $\mathit{BwsA}$, Gent-McWilliams Mesoscale eddy transport coefficient $\mathit{GM}$, Critical Bulk Richardson Number $\mathit{CbrN}$, and Horizontal Viscosity $\mathit{HV}$. Following the methodology of prior work \cite{shi2022vdl}, a 15-model-day simulation was conducted for each parameter configuration, from which a region of interest was extracted centered on the eastern equatorial Pacific cold tongue, bounded by 160°W to 80°E longitude, 26°S to 26°N latitude, and sea level to 200 meters depth, resulting in volumes of size $1536 \times 768 \times 768$ along the longitude, latitude, and depth axes respectively. 
\par
We construct two datasets from this simulation. The first is a direct volume rendering (DVR) dataset, where each ensemble member is volume rendered at a resolution of $512 \times 512$, comprising 70 training and 30 test parameter configurations. The second is an isosurface rendering (IR) dataset, where for each ensemble member we extract isosurfaces of the ocean temperature field at eleven uniformly spaced values in the range [15,25] at $512 \times 512$ resolution. 

\textbf{XCompact3D} \cite{bartholomew2020xcompact3d} is a high-order finite-difference framework for solving the incompressible Navier-Stokes equations on Cartesian meshes, developed for large-scale turbulent flow simulations on high-performance computing platforms. In this study, the domain scientists investigate the Reynolds number parameter $\mathit{Re} \in [1500,2500]$ using a Taylor-Green vortex setup, which is a canonical benchmark in turbulence research.
Each ensemble member produces a volumetric scalar field of the Q-criterion. 
We construct two datasets from this simulation. The first one is a DVR dataset, where each ensemble member's scalar field is rendered at $512 \times 512$ resolution, yielding 100 training and 29 test configurations. The second is an IR dataset, where for each ensemble member we additionally extract isosurfaces at ten uniformly spaced Q-criterion values in the range [-100,-10].

\textbf{CloverLeaf3D}~\cite{biswas2026cloverleaf} is a Lagrangian-Eulerian explicit hydrodynamics mini-application that solves the compressible Euler equations on a 3D structured grid. It models the interaction between a high-density gas region and a surrounding low-density medium, producing a propagating shock front. We study six simulation parameters, and the volume renderings are generated at $256 \times 256$ resolution. We use 200 parameter configurations for training and 40 for testing. 


\begin{table}[htbp]
\centering
\caption{Ensemble simulation datasets: resolutions, number of images, and parameter dimensionalities. For all datasets, the 252 icosphere viewpoints are split evenly, with 126 for training and 126 reserved for testing.}
\label{tab:datasets}
\resizebox{\columnwidth}{!}{
\begin{tabular}{c|c|c|c|c}
Dataset & \makecell{Volume \\ Resolution} & \makecell{\#Views \\ per member} & \makecell{\#Simulation \\ Parameters} & \makecell{Image \\ Resolution} \\
\hline
Nyx & $512 \times 512 \times 512$ & 252 & 3 & $256 \times 256$ \\
MPAS-Ocean & $1536 \times 768 \times 768$ & 252 & 4 & $512 \times 512$ \\
XCompact3D & $256 \times 256 \times 256$ & 252 & 1 & $512 \times 512$ \\
CloverLeaf3D & $128 \times 128 \times 128$ & 252 & 6 & $256 \times 256$ \\
\end{tabular}
}
\end{table}

\subsection{Implementation and Experimental Details}

All experiments are implemented in PyTorch and conducted on a single NVIDIA A100 GPU. In the first training stage, which constructs the canonical field, we use the default learning rates from the original 3DGS implementation for optimizing each Gaussian attribute. 
For adaptive density control, we set the positional gradient threshold to $1 \times 10^{-4}$ and perform pruning, cloning, and splitting of Gaussians every 100 iterations.
In the parameter-conditioned deformation stage, the learning rate of the deformation model is set to $1 \times 10^{-4}$ with an exponential learning rate decay, and the canonical field is jointly finetuned using a 100$\times$ smaller learning rate. 
In both stages, we set the batch size to 1, following the original 3DGS implementation.
For the deformation models, the dimensionality of both spatial and condition feature vectors is set to 128, and the hidden dimension of the MLP layers in the deformation network is set to 512.
We further extend 4DGS to handle high-dimensional parameter settings, denoted as \textit{4DGS-HD} in \cref{sec:comparison_baselines}. For fair comparisons, we use the same canonical Gaussians for both 4DGS-HD and our \ours.

\subsection{Baseline Methods and Evaluation Metrics}

\textbf{Baselines.}
We compare our proposed approach against four baseline methods that support parameter-conditioned visualization synthesis: InSituNet~\cite{he2019insitunet}, VisNeRF~\cite{yao2025visnerf}, K-Planes~\cite{fridovich2023k}, and modified 4DGS~\cite{wu20244d}.
(1) \textit{InSituNet} is a GAN-based surrogate model that synthesizes visualization images directly from simulation parameters and viewing direction, making it a representative image-space baseline for parameter space exploration of ensemble simulations.     
(2) \textit{VisNeRF} extends tensor decomposition-based neural radiance fields with additional parameter-conditioned feature vectors to model volumetric scenes across a continuous simulation parameter space within a single unified model, representing an implicit neural field baseline. 
(3) \textit{K-Planes} factorizes scene representations into a set of 2D feature planes spanning both spatial and parameter dimensions, providing an explicit radiance field baseline that naturally generalizes to high-dimensional ensemble parameter spaces.
(4) \textit{4DGS} represents dynamic scenes using 3D Gaussian primitives coupled with a deformation network that predicts per-Gaussian transformations over time. We modify its deformation network to accept simulation parameters in place of temporal inputs, adapting it to the ensemble setting where appearance and geometry vary across a continuous parameter space rather than along a time axis.
\par
\textbf{Metrics.}
We evaluate the performance of all methods using three visual quality metrics. Peak Signal-to-Noise Ratio (PSNR)~\cite{huynh2008scope} measures the pixel-level reconstruction accuracy between synthesized and ground-truth visualization images, where higher values indicate closer agreement with the reference. Structural Similarity Index Measure (SSIM)~\cite{wang2004image} captures perceptual similarity by jointly assessing luminance, contrast, and structural patterns, providing a more human-aligned measure of image quality than pixel-wise differences alone. Learned Perceptual Image Patch Similarity (LPIPS)~\cite{zhang2018unreasonable} evaluates perceptual similarity using deep feature representations extracted from a pretrained network, capturing high-level appearance differences that PSNR and SSIM may not reflect. 

In addition to image quality, we evaluate the computational efficiency of all methods using three metrics. Model Size (MB) measures the total storage footprint of the trained model, reflecting its practical deployability. Training Time (hr) measures the time required to train each method. Time per Image (s) measures the average inference time required to synthesize an image given a parameter configuration. Together, these metrics complement the image quality measures and allow a more complete assessment of each method's practical utility for parameter space exploration of ensemble simulations.

\subsection{Comparison with Baseline Models}
\label{sec:comparison_baselines}

In this section, we compare our \ours with the four baseline approaches from three perspectives: generalization to the unseen viewpoints (\cref{sec:novel_views}), unseen simulation parameters (\cref{sec:surrogate_prediction}), and unseen isovalues (\cref{sec:isovalue_interpolation}). Finally, we evaluate the robustness of \ours under jointly varying conditions to reflect practical scientific exploration (\cref{sec:joint_conditioning}).


\begin{table}[htbp]
\centering
\caption{Quantitative evaluation of all methods on novel viewpoints across training ensemble members.}
\label{tab:nvs_task}
\setlength{\tabcolsep}{2.2pt}
\small
\begin{tabular}{>{\raggedright\arraybackslash}p{1.5cm}|l|c|c|c|c|c}
Simulation & Metric & InSituNet & K-Planes & ViSNeRF & 4DGS-HD & GS-Surrogate \\
\hline
\multirow{3}{*}{Nyx}
& PSNR$\uparrow$  & 23.07 & 32.62 & 33.56 & 33.47 & \textbf{37.87} \\
& SSIM$\uparrow$  & 0.60  & 0.89  & 0.91  & 0.93  & \textbf{0.97} \\
& LPIPS$\downarrow$ & 0.19 & 0.14  & 0.12  & 0.07  & \textbf{0.03} \\
\hline
\multirow{3}{*}{XCompact3D}
& PSNR$\uparrow$  & 24.86 & 32.27 & 33.10 & 34.44 & \textbf{39.10} \\
& SSIM$\uparrow$  & 0.87 & 0.94 & 0.95 & 0.96 & \textbf{0.98} \\
& LPIPS$\downarrow$ & 0.12 & 0.07 & 0.06 & 0.03 & \textbf{0.01} \\
\hline
\multirow{3}{*}{CloverLeaf3D}
& PSNR$\uparrow$  & 22.34 & 23.32 & \textbf{36.09} & 16.29 & 34.87 \\
& SSIM$\uparrow$  & 0.87 & 0.86 & \textbf{0.97} & 0.82 & \textbf{0.97} \\
& LPIPS$\downarrow$ & 0.14 & 0.27 & 0.04 & 0.31 & \textbf{0.03} \\
\end{tabular}
\end{table}

\subsubsection{Novel View Synthesis}
\label{sec:novel_views}

\begin{figure}[htbp]
    \centering
    \noindent\includegraphics[width=0.99\linewidth]{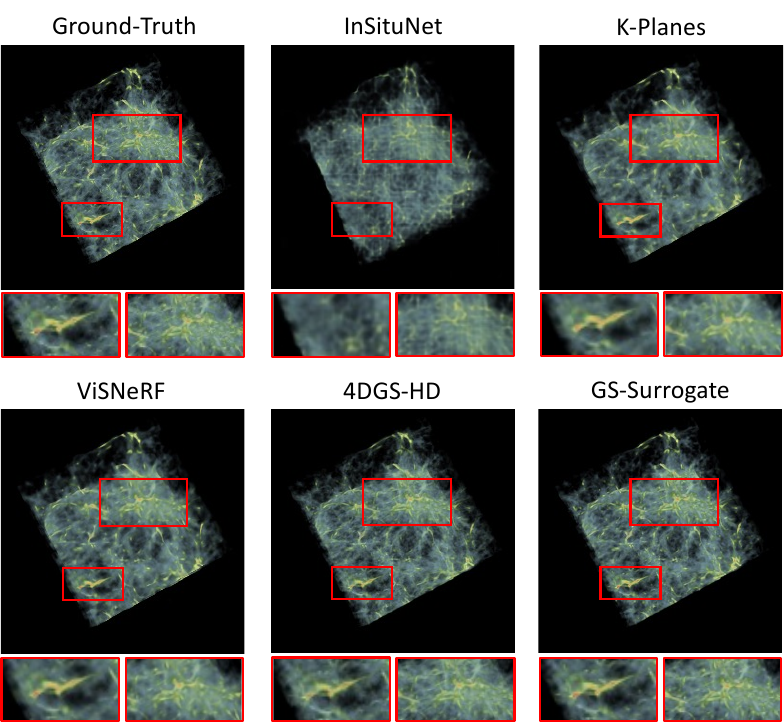}
    \caption{Visual comparison on a training volume-rendering instance from the Nyx dataset under an unseen viewpoint.}
    \label{fig:novel_view_nyx}
\end{figure}

We first evaluate how well each model learns a view-consistent representation from the sparse input images. In this experiment, we measure each model's performance on 126 unseen viewpoints while keeping the simulation parameters the same as those used during training.
As shown in \autoref{tab:nvs_task}, our \ours achieves the best overall performance across all datasets. On CloverLeaf3D, although the PSNR is slightly lower compared to ViSNeRF, our method still maintains competitive perceptual quality with an SSIM of 0.97 and an LPIPS of 0.03.
In contrast, InSituNet consistently performs worse across all three datasets, suggesting that purely image-space surrogate models are less effective in learning view-consistent representations and handling unseen viewpoints.
As further illustrated in \autoref{fig:novel_view_nyx}, under an unseen viewpoint, our method accurately reconstructs the filamentary structures of the Nyx simulation, whereas these fine-scale details are largely missing in the image produced by InSituNet.


\subsubsection{Simulation Parameter Generalization}
\label{sec:surrogate_prediction}

Generalization to unseen simulation parameters is important for parameter-space exploration. Unlike novel view synthesis, where performance is mainly determined by the underlying 3D-aware representation (\ie, NeRF or 3DGS), this setting evaluates how effectively a model can handle variations over a high-dimensional simulation parameter space. \autoref{tab:unseen_simulation_parameters} summarizes the quantitative results on three volume-rendering datasets.
\par
\begin{figure}[htbp]
    \centering
    \noindent\includegraphics[width=1.0\linewidth]{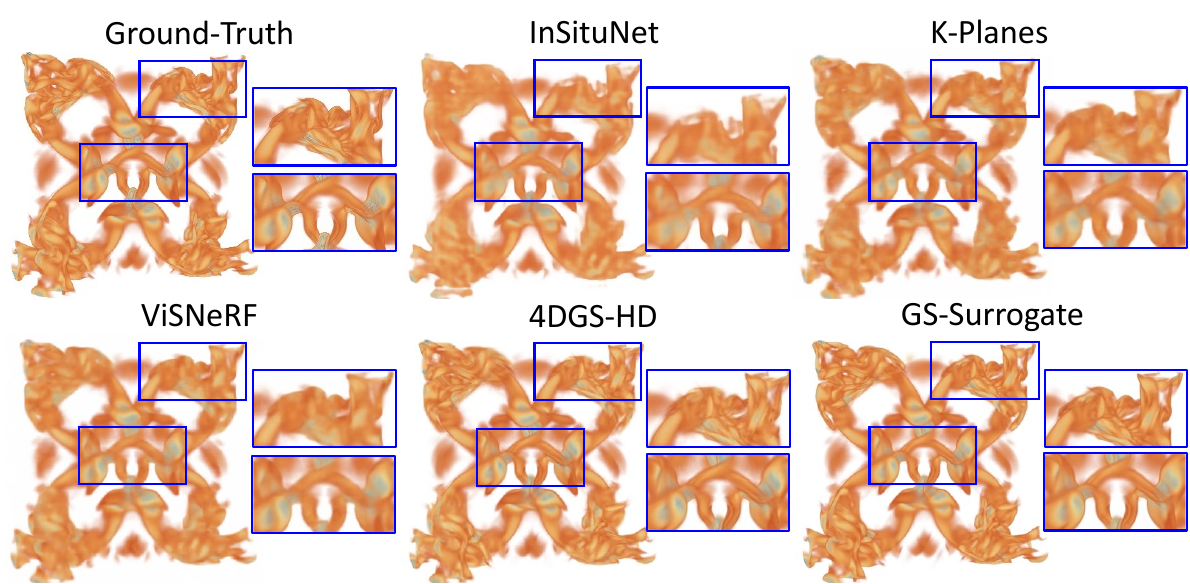}
    \caption{Visual comparison on a representative unseen ensemble member from the XCompact3D dataset.}
    \label{fig:unseen_params_xcompact}
\end{figure}

Our method still achieves the best overall performance on the Nyx and XCompact3D datasets across all three metrics. On CloverLeaf3D, which is the most challenging dataset due to its large variation across a six-dimensional simulation parameter space, our method achieves a slightly lower PSNR than ViSNeRF while maintaining comparable perceptual quality. We further present a visualization result for an unseen ensemble member from the XCompact3D dataset in \autoref{fig:unseen_params_xcompact}. Although all methods are able to recover the overall geometric structure, the zoomed-in views show that our \ours better resolves the fine-grained details than the other approaches.
\par
Moreover, K-Planes shows limited generalization ability as the dimensionality of the simulation parameter space increases. While it performs reasonably well in lower-dimensional settings, \ie, achieving a PSNR of 33.05 dB on XCompact3D with only a 1D simulation parameter, its performance degrades substantially on higher-dimensional datasets.
This issue is particularly evident on CloverLeaf3D, which involves six simulation parameters.
We further compare the top three methods qualitatively on CloverLeaf3D in \autoref{fig:unseen_params_clover}. Overall, InSituNet produces overly smooth predictions and fails to preserve sharp features such as edges and structural transitions. ViSNeRF achieves the highest PSNR on this dataset, likely because NeRF-based methods are more effective for modeling dense volumetric fields such as CloverLeaf3D. However, it also introduces some high-frequency artifacts, especially around the center region of the volume. In contrast, although our method achieves a slightly lower PSNR as some fine-scale structures are still not fully reconstructed, it better preserves the overall perceptual quality without introducing noticeable artifacts.

\begin{table}[htbp]
\centering
\caption{Quantitative evaluation on unseen simulation parameters.}
\label{tab:unseen_simulation_parameters}
\setlength{\tabcolsep}{1.7pt}
\small
\begin{tabular}{>{\raggedright\arraybackslash}p{1.5cm}|l|c|c|c|c|c}
Simulation & Metric & InSituNet & K-Planes & ViSNeRF & 4DGS-HD & GS-Surrogate \\
\hline
\multirow{3}{*}{Nyx}
& PSNR$\uparrow$  & 23.11 & 26.78 & 33.97 & 33.39 & \textbf{38.83} \\
& SSIM$\uparrow$  & 0.61  & 0.80 & 0.92  & 0.93  & \textbf{0.98} \\
& LPIPS$\downarrow$ & 0.20  & 0.27 & 0.12  & 0.07  & \textbf{0.02} \\
\hline
\multirow{3}{*}{XCompact3D}
& PSNR$\uparrow$  & 24.88 & 33.05 & 33.81 & 35.23 & \textbf{41.26} \\
& SSIM$\uparrow$  & 0.86  & 0.95 & 0.95  & 0.97 & \textbf{0.99} \\
& LPIPS$\downarrow$ & 0.12  & 0.07 & 0.06  & 0.02 & \textbf{0.01} \\
\hline
\multirow{3}{*}{CloverLeaf3D}
& PSNR$\uparrow$  & 21.78 & 16.48 & \textbf{32.92} & 16.03 & 30.93 \\
& SSIM$\uparrow$  & 0.87  & 0.79 & \textbf{0.96}  & 0.82  & \textbf{0.96} \\
& LPIPS$\downarrow$ & 0.15  & 0.38 & \textbf{0.05} & 0.31  & \textbf{0.05} \\
\end{tabular}
\end{table}

In addition, we compare the model size, total training time, and per-image inference time across all methods in \autoref{tab:efficiency}. Overall, InSituNet requires the longest training time across all datasets but achieves the fastest inference speed due to its 2D CNN architecture. 
For NeRF-based methods, including ViSNeRF and K-Planes, both training and inference are generally slower than GS-based approaches mainly because ray marching requires dense sampling along each ray. This computational overhead becomes particularly significant for datasets with higher image resolution, \eg, XCompact3D ($512\times512$), since the number of rays scales linearly with the number of pixels.
In contrast, the efficiency of GS-based methods is mainly determined by the number of Gaussian primitives. Although our deformation network is fully MLP-based, it only introduces little computational overhead compared to methods using factorized plane representations, \ie, 4DGS-HD. Meanwhile, our model is less sensitive to the dimensionality of parameters. As a result, our \ours maintains the smallest model size across all datasets while achieving comparable training and inference speeds to 4DGS-HD, with better visual quality.

\begin{table}[htbp]
\centering
\caption{Model size, training time, and per-image inference time for all methods. For GS-based methods, the model size includes both the Gaussian primitives and the parameters of the deformation network.}
\label{tab:efficiency}
\setlength{\tabcolsep}{1.8pt}
\small
\begin{tabular}{l|ccc|ccc|ccc}
 & \multicolumn{3}{c|}{Nyx} & \multicolumn{3}{c|}{XCompact3D} & \multicolumn{3}{c}{CloverLeaf3D} \\
 \hline
 & Size & Train & Test & Size & Train & Test & Size & Train & Test \\
 & (MB) & (hr) & (s/img) & (MB) & (hr) & (s/img) & (MB) & (hr) & (s/img) \\
\hline
InSituNet    & 232.64 & 27.30 & \textbf{0.03} & 232.93 & 45.47 & \textbf{0.05} & 232.65 & 38.70 & \textbf{0.03} \\
K-Planes     & 76.95 & 7.25 & 0.27 & 69.73 & 16.67 & 0.72 & 88.26 & 14.50 & 0.33 \\
ViSNeRF      & 66.43 & 5.05 & 0.18 & 66.43 & 8.65  & 0.73 & 66.43 & \textbf{6.13} & 0.37 \\
4DGS-HD      & \underline{50.75} & \textbf{3.26} & \underline{0.06} & \textbf{43.70} & \textbf{3.05} & \underline{0.06} & \underline{50.53}  & 8.18 & \underline{0.05} \\
GS-Surrogate & \textbf{46.19} & \underline{3.28} & \underline{0.06} & \underline{45.50} & \underline{3.08} & \underline{0.06} & \textbf{38.46 } & \underline{7.87} & \underline{0.05} \\
\end{tabular}
\end{table}
\begin{figure}[b!]
    \centering
    \noindent\includegraphics[width=0.98\linewidth]{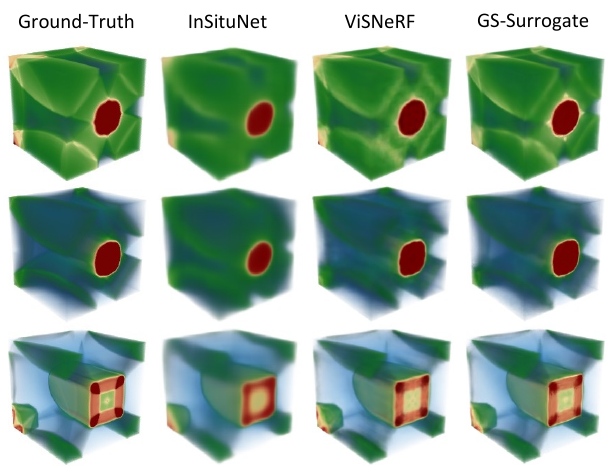}
    \caption{Comparative results of the top three methods on three unseen ensemble members from the CloverLeaf3D dataset. CloverLeaf3D is the most challenging dataset, as its simulation parameters vary substantially across a 6D space.}
    \label{fig:unseen_params_clover}
\end{figure}

\subsubsection{Isovalue Generalization}
\label{sec:isovalue_interpolation}

In this isosurface extraction task, we evaluate the generalization ability of each model to the unseen isovalues. Specifically, on the MPAS-Ocean dataset, we consider a temperature range from 15 to 25, using eight isovalues for training and three for testing.
To ensure a fair comparison, both 4DGS-HD and \ours follows the training pipeline described in \cref{sec:method}, where the full ocean volume is first constructed and then deformed into different isosurfaces given the training isovalues. This task is particularly challenging because the model must generalize across different isovalues each with different geometric structures. Especially for 4DGS-HD and our \ours, the model needs to learn how to deform a volumetric representation into corresponding surfaces. 
\par
\begin{figure}[htbp]
    \centering
    \noindent\includegraphics[width=1.0\linewidth]{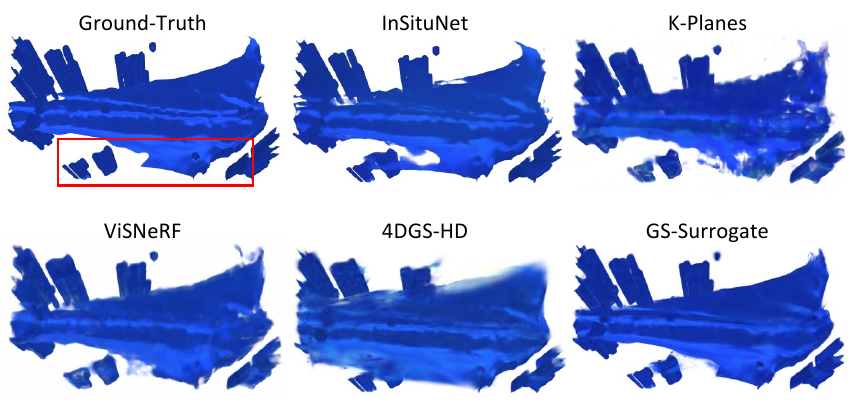}
    \caption{Qualitative comparison on the MPAS-Ocean dataset under an unseen isovalue of temperature value 16. The most challenging region to reconstruct is highlighted in red.}
    \label{fig:unseen_isovalues}
\end{figure}

\autoref{tab:unseen_isovalues} summarizes the quantitative results for all methods. 
Among the baseline approaches, InSituNet, K-Planes, and 4DGS-HD achieve similar PSNR values. Notably, InSituNet obtains a better LPIPS score of 0.08, indicating its strength in preserving the overall visual appearance. This is also reflected in the qualitative results in \autoref{fig:unseen_isovalues}. 
ViSNeRF achieves a higher PSNR compared to these three baselines, but its rendered result still presents several under-reconstructed regions.
In contrast, \ours achieves the best performance across all three metrics. Although some fine-scale details still require improvement, our method reconstructs the isosurface geometry more faithfully.

\begin{table}[htbp]
\centering
\caption{Performance on unseen isovalues for isosurface extraction on the MPAS-Ocean dataset.}
\label{tab:unseen_isovalues}
\setlength{\tabcolsep}{4pt}
\small
\begin{tabular}{l|c|c|c|c|c}
Metric & InSituNet & K-Planes & ViSNeRF & 4DGS-HD & GS-Surrogate \\
\hline
PSNR$\uparrow$  & 20.96 & 21.71 & \underline{23.57} & 20.97 & \textbf{26.12} \\
SSIM$\uparrow$  & \underline{0.91} & 0.87 & 0.90 & 0.90 & \textbf{0.94} \\
LPIPS$\downarrow$ & \underline{0.08} & 0.15 & 0.11 & 0.13 & \textbf{0.06} \\
\end{tabular}
\end{table}

\subsubsection{Joint Parameter Conditioning}
\label{sec:joint_conditioning}

In practice, scientists often need to explore the simulations across arbitrary viewpoints and parameter settings. Therefore, beyond evaluating each condition independently, we further evaluate \ours under the joint generalization setting of unseen viewpoints and unseen simulation parameters across three volume-rendering and two isosurface-extraction tasks. As shown in \autoref{tab:unseen_params_unseen_views}, our method maintains stable performance in this more challenging setting. 

\begin{table}[htbp]
\centering
\caption{Evaluation of \ours under unseen viewpoints and simulation parameters across three volume rendering tasks and two isosurface extraction tasks.}
\label{tab:unseen_params_unseen_views}
\setlength{\tabcolsep}{3pt}
\small
\begin{tabular}{l|c|c|c|c|c}
Dataset & Nyx & XCompact3D & CloverLeaf3D & MPAS-Ocean & XCompact3D \\
\hline
Resolution & $256\times256$ & $512\times512$ & $256\times256$ & $512\times512$ & $512\times512$ \\
Task & DVR & DVR & DVR & IR & IR \\
\hline
PSNR$\uparrow$  & 37.96 & 39.98 & 30.64 & 30.05 & 30.65 \\
SSIM$\uparrow$   & 0.97  & 0.98  & 0.96  & 0.95  & 0.95 \\
LPIPS$\downarrow$ & 0.02  & 0.01  & 0.05  & 0.05  & 0.05 \\
\end{tabular}
\end{table}
\begin{figure}[htbp]
    \centering
    \noindent\includegraphics[width=1.0\linewidth]{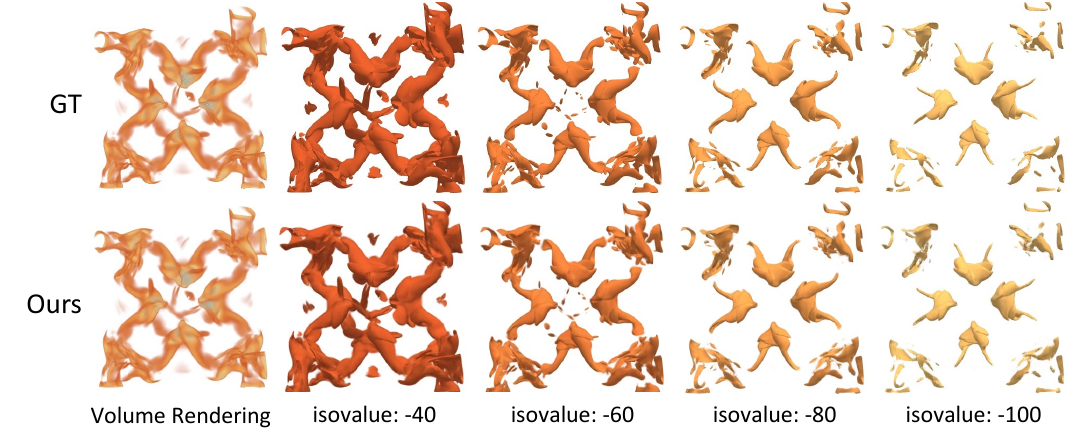}
    \caption{Visualization results of four different isosurfaces from the XCompact3D dataset under both an unseen viewpoint and unseen simulation parameter.}
    \label{fig:xcompact_test_unseen}
\end{figure}

\begin{figure}[htbp]
    \centering
    \noindent\includegraphics[width=0.88\linewidth]{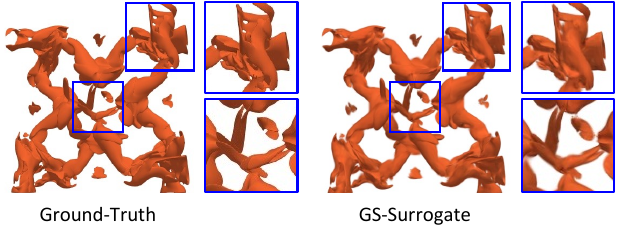}
    \caption{Zoomed-in views highlighting the visual differences between the ground truth and \ours in fine-scale structures.}
    \label{fig:compare_isosurfaces}
\end{figure}

We further present an unseen ensemble member from the XCompact3D isosurface extraction task in \autoref{fig:xcompact_test_unseen}. During training, the deformation model $F_{\text{sim}}$ first learns to generalize across the simulation parameter space, and $F_{\text{vis}}$ then deforms the volumetric representation into isosurfaces conditioned on specific isovalues. 
Compared to the ground truth, our method produces high-quality isosurfaces that closely match both the overall geometry and surface appearance. In particular, the model accurately captures the gradual structural changes across different isovalues. 
However, as shown in the zoomed-in view in \autoref{fig:compare_isosurfaces}, our predictions are still slightly blurred in the fine-scale surface regions. This indicates that modeling the high-frequency geometric details under multiple varying parameters (\ie, viewpoints, simulation parameters, and isovalues) remains challenging and could be further improved in future work.

\section{Parameter Space Exploration with GS-Surrogate}
\label{sec:exploration}

\subsection{Visual Interface}

\begin{figure}[htbp]
    \centering
    \includegraphics[width=0.98\linewidth]{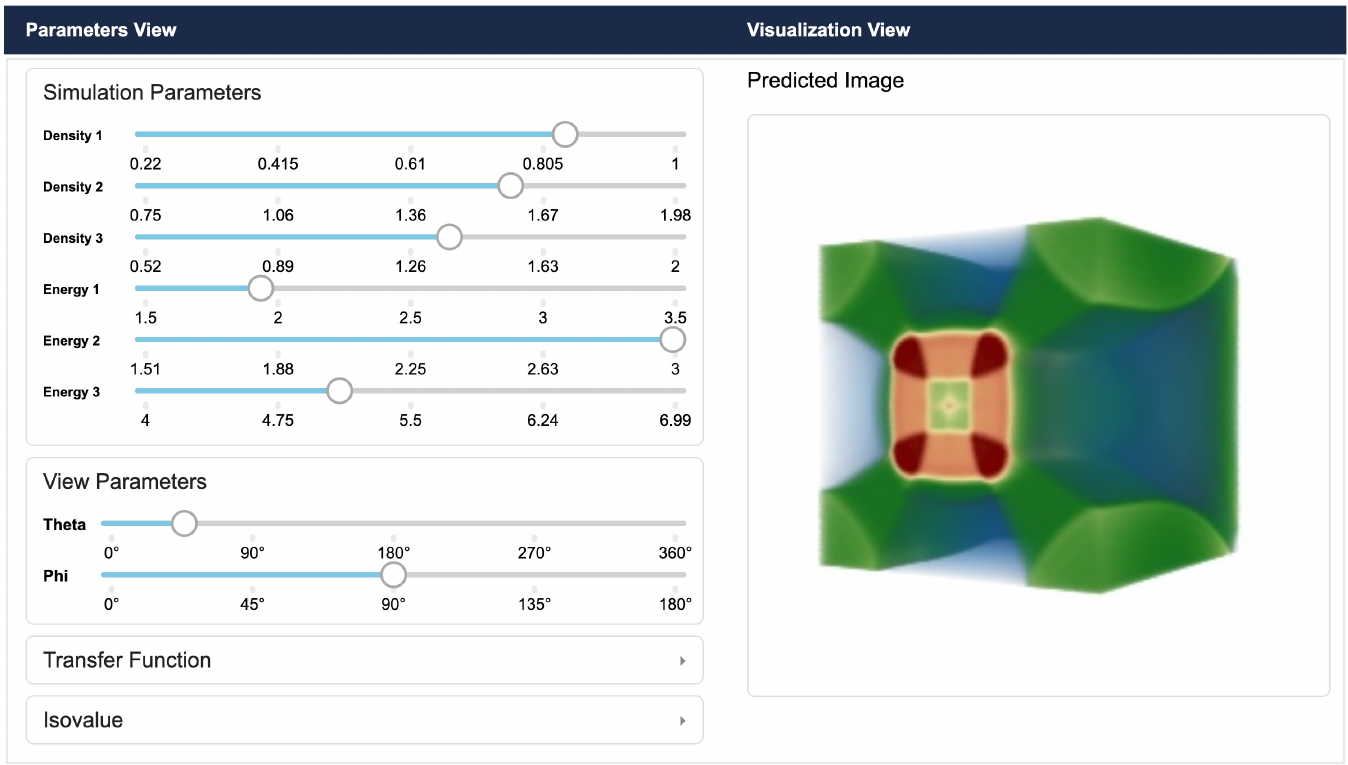}
    \caption{Interface for interactive parameter space exploration with GS-Surrogate.}
    \label{fig:Interface}
\end{figure}

GS-Surrogate allows interactive post-hoc exploration of ensemble simulations through a unified visual interface (see \autoref{fig:Interface}). The interface organizes parameters into three complementary groups: simulation parameters, view parameters, and visualization parameters. The visualization controls are provided through two separate panels, allowing users to adjust transfer functions or select isovalues as needed.
Given any combination of these inputs, the trained GS-Surrogate performs forward inference by deforming the canonical Gaussian field and directly producing the corresponding visualization, allowing users to continuously explore the simulation parameter space.
The transfer function editor provides fine-grained control over opacity mappings, supporting the highlighting of localized features and facilitating detailed visual analysis.

\subsection{Case Study with the Nyx Simulation}

\begin{figure}[htbp]
    \centering
    \includegraphics[width=0.99\linewidth]{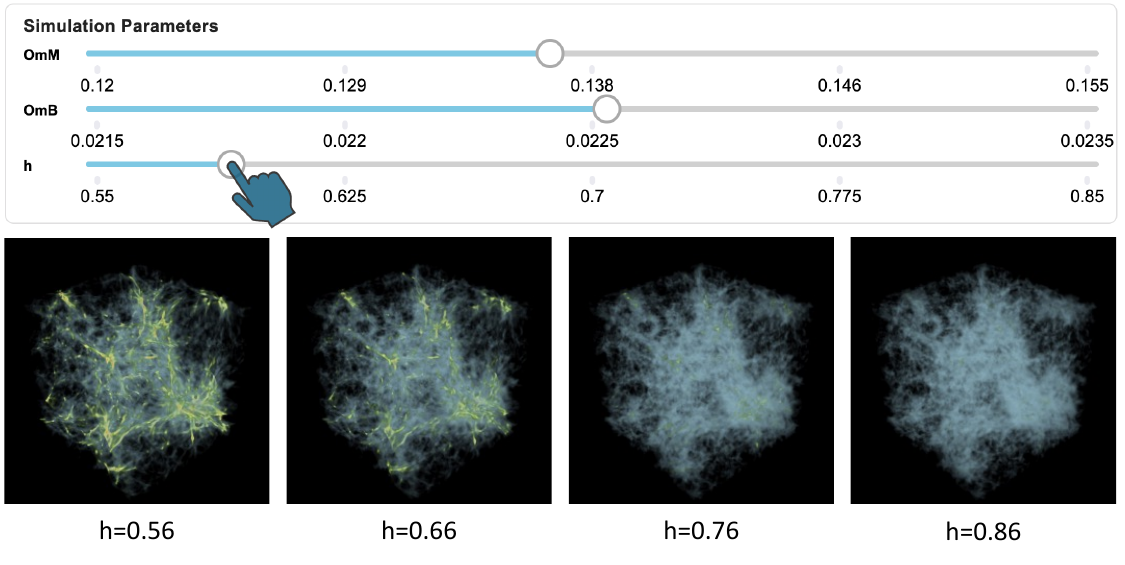}
    \caption{Predicted volume renderings varying $\mathit{h} \in \{0.56,0.66,0.76,0.86\}$ on the Nyx dataset, with $\mathit{OmM}$ and $\mathit{OmB}$ fixed.}
    \label{fig:explore_nyx_params}
\end{figure}

To demonstrate the utility of GS-Surrogate for scientific discovery, this case study focused on the Nyx cosmological dataset. As shown in \autoref{fig:explore_nyx_params} the interactive system allows scientists to navigate the multi-dimensional parameter space of the dataset. Unlike traditional workflows that require expensive on-the-fly rendering or I/O-heavy data loading, GS-Surrogate provides instantaneous visual feedback. The exploration process begins with scientists selecting the initial parameters within valid ranges, serving as an entry point for further investigation. From this starting point, scientists can systematically refine their exploration based on prior knowledge and observed visual patterns. 
By interactively adjusting the three simulation parameters (\ie, $\mathit{OmM}$, $\mathit{OmB}$, and $\mathit{h}$), domain scientists can directly examine how variations in these parameters affect the spatial distribution and density structures of the cosmological field.
In addition to parameter selection, the system supports flexible viewpoint control, allowing scientists to inspect structures from multiple perspectives.

Another key component of the exploration workflow is the transfer function editor, which provides fine-grained control over how scalar values are mapped to visual appearance. 
In our interface, the transfer function is defined through four interactive control points that specify a piecewise linear mapping between scalar values and opacity. 
To ensure controllable and meaningful exploration, we fix the endpoint opacities of the transfer function: the first control point, corresponding to the lowest scalar values, is set to zero opacity, whereas the last control point, corresponding to the highest scalar values, is fixed at full opacity. This design preserves the visibility of high-frequency, high-density structures while preventing low-value noise from dominating the visualization. 
The two intermediate control points, $c1$ and $c2$, are exposed for user interaction and provide targeted control over different regions of the scalar field. $c1$ primarily governs the visibility of low to mid-density gaseous structures, allowing scientists to selectively enhance or suppress those regions. $c2$ focuses on higher scalar ranges, allowing scientists to refine the visibility of denser and more detailed structures. 
Together, by adjusting these control points, scientists can reshape the opacity curve and balance the visibility of faint filamentary features against dominant high-density regions. This is especially important for cosmological data, where the scalar field spans a broad range of values and subtle structures can be easily obscured. By interactively adjusting the transfer function, scientists can emphasize density intervals of interest and perform more precise and flexible analysis.
\par
In this example, the scientist first fixes the viewpoint and then sweeps through the range of $\mathit{h}$. As shown in \autoref{fig:explore_nyx_params}, the GS-Surrogate predictions reveal a clear and physically consistent trend: lower values of $\mathit{h}$ lead to more spatially concentrated matter distributions with higher local density contrast, whereas higher values produce more diffuse structures. Once a simulation configuration of interest is identified, scientists can further refine their analysis by interactively adjusting the opacity mapping to reveal different structural features.

\begin{figure}[htbp]
    \centering
    \includegraphics[width=0.99\linewidth]{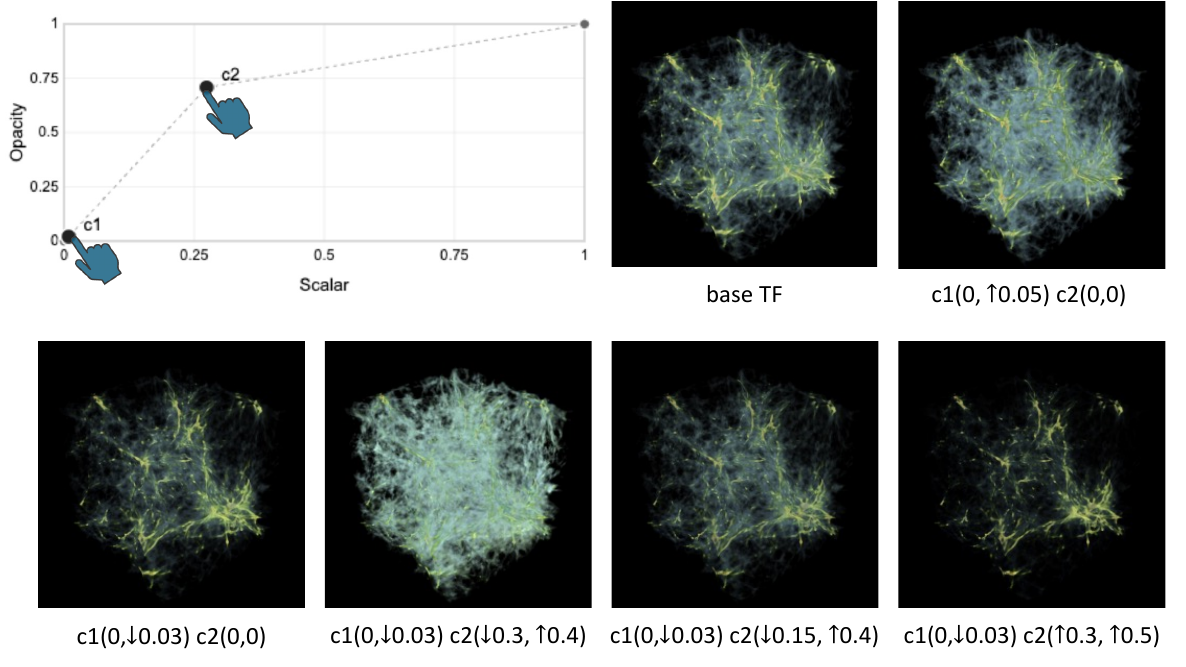}
    \caption{TF space exploration on the Nyx dataset with fixed simulation parameters. Starting from the base TF, the two interior control points $c1$ and $c2$ are displaced in scalar and opacity to selectively reveal or suppress features. }
    \label{fig:explore_nyx_tfs}
\end{figure}

\autoref{fig:explore_nyx_tfs} demonstrates transfer function exploration by varying the two control points, $c1$ and $c2$, while keeping the simulation parameters fixed.
To facilitate visual comparison, each rendered image is shown together with the absolute difference from the base TF. Starting from the base TF, adjusting $c1$ upward in opacity by 0.05 (\ie, $c1(0, \uparrow 0.05)$) increases the visual contribution of lower-density matter, bringing previously suppressed low-frequency structures into view. Conversely, reducing the opacity of $c1$ by 0.03 (\ie, $c1(0, \downarrow 0.03)$) suppresses these lower-density regions, sharpening the visual emphasis on higher-density structures. The subsequent variations jointly modify both $c1$ and $c2$: displacing $c2$ in both scalar and opacity dimensions (\ie, $c2(\downarrow 0.3, \uparrow 0.4)$, $c2(\downarrow 0.15, \uparrow 0.4)$, $c2(\uparrow 0.3, \uparrow 0.5)$) reshapes the opacity curve in the mid-to-high scalar range, selectively revealing or suppressing different density regimes of the dark matter distribution. 
The absolute difference images confirm that these modifications produce spatially structured, non-trivial changes in the rendered output, validating that GS-Surrogate's deformation model accurately captures opacity-driven appearance changes. Together, \autoref{fig:explore_nyx_params} and \autoref{fig:explore_nyx_tfs} demonstrate that GS-Surrogate supports flexible and scientifically meaningful post-hoc exploration across both the simulation and visualization parameter spaces.

\section{Discussion and Future Work}

In this section, we discuss our current limitations and several directions for future improvement. 
First, for scientific fields with highly dense features and large variations across parameter space, such as the CloverLeaf3D dataset, the current method still requires a larger number of Gaussian primitives or more effective deformation approaches to better handle such variations in dense regions. 
To better capture localized small-scale structures, one possible direction is to allocate more Gaussians to increase the representational capacity. However, this would come at the cost of substantially higher memory usage. Another direction is to augment the Gaussian primitives with additional texture representations to better capture the high-frequency details~\cite{chao2025textured}. While such advanced representations may alleviate the memory overhead and improve reconstruction quality, they often introduce additional training complexity.
\par
Second, there is still room to further improve the performance of our current framework on the isosurface extraction task.
Compared to volume rendering, this task requires the model not only to resolve fine-scale structures but also to deform a volumetric representation into surfaces that share entirely different topological structures with the original volume. Specifically, volume rendering relies on semi-transparent Gaussians distributed across the entire field, whereas isosurface extraction requires more opaque Gaussians concentrated around the target surface. As a result, learning a fixed set of Gaussian representations that can be efficiently adapted across different visualization tasks is still challenging. For further work, we could further improve the effectiveness of visualization-conditioned deformation for this setting. Moreover, the current transfer function editing is limited to opacity mapping and could be further extended to support more flexible changes, such as different color mappings.
\par
Finally, another promising direction for future work is to extend \ours to support bidirectional prediction. Currently, our framework synthesizes visualization results from given simulation parameters and visualization settings, which only operates in the forward direction. However, in scientific analysis, the scientists are often interested in knowing what parameter configurations could produce a target structure of interest. Since both the model and rendering pipeline are fully differentiable, this reverse prediction could be achieved by optimizing the input parameters through backpropagation from the target image.


\section{Conclusion}

In this paper, we present \ours, a deformable 3D Gaussian Splatting-based visualization surrogate for supporting interactive post-hoc exploration of ensemble simulations. Compared to the prior work, which either relies on synthesizing the rendered images in the 2D image space or models all parameter-driven variation within a single implicit neural radiance field, our method leverages the parametrized Gaussian primitives and explicitly decouples the learning into two steps. From the canonical Gaussians, \ours first learns to adapt to the structural changes across the simulation parameter space and then further deforms the Gaussian primitives for the target visualization task. Our modularized framework facilitates more effective and controllable exploration across different visualization tasks, \eg, isosurface extraction and transfer function editing.

\section{Acknowledgments}

This work was supported by the U.S. Department of Energy, Office of Science, Office of Advanced Scientific Computing Research's Computer Science Competitive Portfolios program under Contract No. DE-AC05-00OR22725.
This research used resources of the Oak Ridge Leadership Computing Facility at the Oak Ridge National Laboratory, which is supported by the Advanced Scientific Computing Research programs in the Office of Science of the U.S. Department of Energy under Contract No. DE-AC05-00OR22725.




\bibliographystyle{abbrv-doi-hyperref}

\bibliography{template}

@article{ma2009situ,
  title={In situ visualization at extreme scale: Challenges and opportunities},
  author={Ma, Kwan-Liu},
  journal={IEEE Computer Graphics and Applications},
  volume={29},
  number={6},
  pages={14--19},
  year={2009},
  publisher={IEEE},
  doi = {10.1109/MCG.2009.120}
}

@inproceedings{bauer2016situ,
  title={In situ methods, infrastructures, and applications on high performance computing platforms},
  author={Bauer, Andrew C and Abbasi, Hasan and Ahrens, James and Childs, Hank and Geveci, Berk and Klasky, Scott and Moreland, Kenneth and O'Leary, Patrick and Vishwanath, Venkatram and Whitlock, Brad and others},
  booktitle={Computer Graphics Forum},
  volume={35},
  number={3},
  pages={577--597},
  year={2016},
  organization={Wiley Online Library},
  doi = {https://doi.org/10.1111/cgf.12930}
}

@article{han2022coordnet,
  title={Coordnet: Data generation and visualization generation for time-varying volumes via a coordinate-based neural network},
  author={Han, Jun and Wang, Chaoli},
  journal={IEEE Transactions on Visualization and Computer Graphics},
  volume={29},
  number={12},
  pages={4951--4963},
  year={2022},
  publisher={IEEE},
  doi = {10.1109/TVCG.2022.3197203}
}

@inproceedings{chen2022tensorf,
  title={Tensorf: Tensorial radiance fields},
  author={Chen, Anpei and Xu, Zexiang and Geiger, Andreas and Yu, Jingyi and Su, Hao},
  booktitle={European conference on computer vision},
  pages={333--350},
  year={2022},
  organization={Springer},
  doi={
https://doi.org/10.48550/arXiv.2203.09517}
}

@article{kerbl20233d,
  title={3d gaussian splatting for real-time radiance field rendering.},
  author={Kerbl, Bernhard and Kopanas, Georgios and Leimk{\"u}hler, Thomas and Drettakis, George and others},
  journal={ACM Trans. Graph.},
  volume={42},
  number={4},
  pages={139--1},
  year={2023},
  doi={
https://doi.org/10.48550/arXiv.2308.04079}
}

@inproceedings{fridovich2022plenoxels,
  title={Plenoxels: Radiance fields without neural networks},
  author={Fridovich-Keil, Sara and Yu, Alex and Tancik, Matthew and Chen, Qinhong and Recht, Benjamin and Kanazawa, Angjoo},
  booktitle={Proceedings of the IEEE/CVF conference on computer vision and pattern recognition},
  pages={5501--5510},
  year={2022},
  doi={
https://doi.org/10.48550/arXiv.2112.05131
}
}

@article{mildenhall2021nerf,
  title={Nerf: Representing scenes as neural radiance fields for view synthesis},
  author={Mildenhall, Ben and Srinivasan, Pratul P and Tancik, Matthew and Barron, Jonathan T and Ramamoorthi, Ravi and Ng, Ren},
  journal={Communications of the ACM},
  volume={65},
  number={1},
  pages={99--106},
  year={2021},
  publisher={ACM New York, NY, USA},
  doi={
https://doi.org/10.48550/arXiv.2003.08934
}
}

@inproceedings{ahrens2014image,
  title={An image-based approach to extreme scale in situ visualization and analysis},
  author={Ahrens, James and Jourdain, S{\'e}bastien and O'Leary, Patrick and Patchett, John and Rogers, David H and Petersen, Mark},
  booktitle={SC'14: Proceedings of the International Conference for High Performance Computing, Networking, Storage and Analysis},
  pages={424--434},
  year={2014},
  organization={IEEE},
  doi={10.1109/SC.2014.40}
}

@inproceedings{biedert2015contour,
  title={Contour Tree Depth Images For Large Data Visualization.},
  author={Biedert, Tim and Garth, Christoph},
  booktitle={EGPGV@ EuroVis},
  pages={77--86},
  year={2015},
  doi={10.2312/pgv.20151158}
}

@inproceedings{frey2013explorable,
  title={Explorable volumetric depth images from raycasting},
  author={Frey, Steffen and Sadlo, Filip and Ertl, Thomas},
  booktitle={2013 XXVI Conference on Graphics, Patterns and Images},
  pages={123--130},
  year={2013},
  organization={IEEE},
  doi={10.1109/SIBGRAPI.2013.26}
}

@article{he2019insitunet,
  title={InSituNet: Deep image synthesis for parameter space exploration of ensemble simulations},
  author={He, Wenbin and Wang, Junpeng and Guo, Hanqi and Wang, Ko-Chih and Shen, Han-Wei and Raj, Mukund and Nashed, Youssef SG and Peterka, Tom},
  journal={IEEE transactions on visualization and computer graphics},
  volume={26},
  number={1},
  pages={23--33},
  year={2019},
  publisher={IEEE},
  doi = {https://doi.org/10.1109/TVCG.2019.2934312}
}

@article{berger2018generative,
  title={A generative model for volume rendering},
  author={Berger, Matthew and Li, Jixian and Levine, Joshua A},
  journal={IEEE transactions on visualization and computer graphics},
  volume={25},
  number={4},
  pages={1636--1650},
  year={2018},
  publisher={IEEE},
  doi={
https://doi.org/10.1109/TVCG.2018.2816059
}
}

@article{shi2022vdl,
  title={VDL-Surrogate: A view-dependent latent-based model for parameter space exploration of ensemble simulations},
  author={Shi, Neng and Xu, Jiayi and Li, Haoyu and Guo, Hanqi and Woodring, Jonathan and Shen, Han-Wei},
  journal={IEEE Transactions on Visualization and Computer Graphics},
  volume={29},
  number={1},
  pages={820--830},
  year={2022},
  publisher={IEEE},
  doi={
https://doi.org/10.48550/arXiv.2207.13091}
}

@inproceedings{yao2025visnerf,
  title={ViSNeRF: Efficient multidimensional neural radiance field representation for visualization synthesis of dynamic volumetric scenes},
  author={Yao, Siyuan and Lu, Yunfei and Wang, Chaoli},
  booktitle={2025 IEEE 18th Pacific Visualization Conference (PacificVis)},
  pages={235--245},
  year={2025},
  organization={IEEE},
  doi={
https://doi.org/10.48550/arXiv.2502.16731}
}

@article{shi2022gnn,
  title={Gnn-surrogate: A hierarchical and adaptive graph neural network for parameter space exploration of unstructured-mesh ocean simulations},
  author={Shi, Neng and Xu, Jiayi and Wurster, Skylar W and Guo, Hanqi and Woodring, Jonathan and Van Roekel, Luke P and Shen, Han-Wei},
  journal={IEEE Transactions on Visualization and Computer Graphics},
  volume={28},
  number={6},
  pages={2301--2313},
  year={2022},
  publisher={IEEE},
  doi={10.1109/TVCG.2022.3165345}
}

@article{obermaier2015visual,
  title={Visual trends analysis in time-varying ensembles},
  author={Obermaier, Harald and Bensema, Kevin and Joy, Kenneth I},
  journal={IEEE transactions on visualization and computer graphics},
  volume={22},
  number={10},
  pages={2331--2342},
  year={2015},
  publisher={IEEE},
  doi={10.1109/TVCG.2015.2507592}
}

@article{matkovic2009interactive,
  title={Interactive visual analysis of complex scientific data as families of data surfaces},
  author={Matkovic, Kresimir and Gracanin, Denis and Klarin, Borislav and Hauser, Helwig},
  journal={IEEE Transactions on Visualization and Computer Graphics},
  volume={15},
  number={6},
  pages={1351--1358},
  year={2009},
  publisher={IEEE},
  doi={10.1109/TVCG.2009.155}
}

@article{orban2018drag,
  title={Drag and track: A direct manipulation interface for contextualizing data instances within a continuous parameter space},
  author={Orban, Daniel and Keefe, Daniel F and Biswas, Ayan and Ahrens, James and Rogers, David},
  journal={IEEE transactions on visualization and computer graphics},
  volume={25},
  number={1},
  pages={256--266},
  year={2018},
  publisher={IEEE},
  doi={https://doi.org/10.1109/tvcg.2018.2865051}
}

@article{bruckner2010result,
  title={Result-driven exploration of simulation parameter spaces for visual effects design},
  author={Bruckner, Stefan and M{\"o}ller, Torsten},
  journal={IEEE Transactions on Visualization and Computer Graphics},
  volume={16},
  number={6},
  pages={1468--1476},
  year={2010},
  publisher={IEEE},
  doi={https://doi.org/10.1109/TVCG.2010.190}
}

@article{chen2015uncertainty,
  title={Uncertainty-aware multidimensional ensemble data visualization and exploration},
  author={Chen, Haidong and Zhang, Song and Chen, Wei and Mei, Honghui and Zhang, Jiawei and Mercer, Andrew and Liang, Ronghua and Qu, Huamin},
  journal={IEEE transactions on visualization and computer graphics},
  volume={21},
  number={9},
  pages={1072--1086},
  year={2015},
  publisher={IEEE},
  doi={10.1109/TVCG.2015.2410278}
}

@article{wang2016multi,
  title={Multi-resolution climate ensemble parameter analysis with nested parallel coordinates plots},
  author={Wang, Junpeng and Liu, Xiaotong and Shen, Han-Wei and Lin, Guang},
  journal={IEEE transactions on visualization and computer graphics},
  volume={23},
  number={1},
  pages={81--90},
  year={2016},
  publisher={IEEE},
  doi={10.1109/TVCG.2016.2598830}
}

@inproceedings{bock2015visual,
  title={Visual verification of space weather ensemble simulations},
  author={Bock, Alexander and Pembroke, Asher and Mays, M Leila and Rastaetter, Lutz and Ropinski, Timo and Ynnerman, Anders},
  booktitle={2015 IEEE Scientific Visualization Conference (SciVis)},
  pages={17--24},
  year={2015},
  organization={IEEE},
  doi={10.1109/SciVis.2015.7429487}
}

@article{poco2014visual,
  title={Visual reconciliation of alternative similarity spaces in climate modeling},
  author={Poco, Jorge and Dasgupta, Aritra and Wei, Yaxing and Hargrove, William and Schwalm, Christopher R and Huntzinger, Deborah N and Cook, Robert and Bertini, Enrico and Silva, Claudio T},
  journal={IEEE transactions on visualization and computer graphics},
  volume={20},
  number={12},
  pages={1923--1932},
  year={2014},
  publisher={IEEE},
  doi={10.1109/TVCG.2014.2346755}
}

@article{almgren2013nyx,
  title={Nyx: A massively parallel amr code for computational cosmology},
  author={Almgren, Ann S and Bell, John B and Lijewski, Mike J and Luki{\'c}, Zarija and Van Andel, Ethan},
  journal={The Astrophysical Journal},
  volume={765},
  number={1},
  pages={39},
  year={2013},
  publisher={The American Astronomical Society},
  doi={https://doi.org/10.1088/0004-637X/765/1/39}
}

@techreport{biswas2026cloverleaf,
  title={Cloverleaf Data Artifacts for ArtIMis LDRD},
  author={Biswas, Ayan and Turton, Terece},
  year={2026},
  institution={Los Alamos National Laboratory (LANL)},
  doi = {10.25583/3022785}
}

@article{bartholomew2020xcompact3d,
  title={Xcompact3D: An open-source framework for solving turbulence problems on a Cartesian mesh},
  author={Bartholomew, Paul and Deskos, Georgios and Frantz, Ricardo AS and Schuch, Felipe N and Lamballais, Eric and Laizet, Sylvain},
  journal={SoftwareX},
  volume={12},
  pages={100550},
  year={2020},
  publisher={Elsevier},
  doi={10.1016/j.softx.2020.100550}
}

@article{ringler2013multi,
  title={A multi-resolution approach to global ocean modeling},
  author={Ringler, Todd and Petersen, Mark and Higdon, Robert L and Jacobsen, Doug and Jones, Philip W and Maltrud, Mathew},
  journal={Ocean Modelling},
  volume={69},
  pages={211--232},
  year={2013},
  publisher={Elsevier},
  doi={https://doi.org/10.1016/j.ocemod.2013.04.010}
}

@inproceedings{fridovich2023k,
  title={K-planes: Explicit radiance fields in space, time, and appearance},
  author={Fridovich-Keil, Sara and Meanti, Giacomo and Warburg, Frederik Rahb{\ae}k and Recht, Benjamin and Kanazawa, Angjoo},
  booktitle={Proceedings of the IEEE/CVF conference on computer vision and pattern recognition},
  pages={12479--12488},
  year={2023},
  doi={https://doi.org/10.48550/arXiv.2301.10241}
}

@inproceedings{chao2025textured,
  title={Textured gaussians for enhanced 3d scene appearance modeling},
  author={Chao, Brian and Tseng, Hung-Yu and Porzi, Lorenzo and Gao, Chen and Li, Tuotuo and Li, Qinbo and Saraf, Ayush and Huang, Jia-Bin and Kopf, Johannes and Wetzstein, Gordon and others},
  booktitle={Proceedings of the Computer Vision and Pattern Recognition Conference},
  pages={8964--8974},
  year={2025},
  doi = {https://doi.org/10.48550/arXiv.2411.18625}
}

@article{wang2004image,
  title={Image quality assessment: from error visibility to structural similarity},
  author={Wang, Zhou and Bovik, Alan C and Sheikh, Hamid R and Simoncelli, Eero P},
  journal={IEEE transactions on image processing},
  volume={13},
  number={4},
  pages={600--612},
  year={2004},
  publisher={IEEE},
  doi={10.1109/TIP.2003.819861}
}

@article{huynh2008scope,
  title={Scope of validity of PSNR in image/video quality assessment},
  author={Huynh-Thu, Quan and Ghanbari, Mohammed},
  journal={Electronics letters},
  volume={44},
  number={13},
  pages={800--801},
  year={2008},
  publisher={IET},
  doi={https://doi.org/10.1049/el:20080522}
}

@inproceedings{zhang2018unreasonable,
  title={The unreasonable effectiveness of deep features as a perceptual metric},
  author={Zhang, Richard and Isola, Phillip and Efros, Alexei A and Shechtman, Eli and Wang, Oliver},
  booktitle={Proceedings of the IEEE conference on computer vision and pattern recognition},
  pages={586--595},
  year={2018},
  doi={https://doi.org/10.48550/arXiv.1801.03924}
}

@inproceedings{cao2023hexplane,
  title={Hexplane: A fast representation for dynamic scenes},
  author={Cao, Ang and Johnson, Justin},
  booktitle={Proceedings of the IEEE/CVF Conference on Computer Vision and Pattern Recognition},
  pages={130--141},
  year={2023},
  doi = {https://doi.org/10.48550/arXiv.2301.09632}
}

@inproceedings{wu20244d,
  title={4d gaussian splatting for real-time dynamic scene rendering},
  author={Wu, Guanjun and Yi, Taoran and Fang, Jiemin and Xie, Lingxi and Zhang, Xiaopeng and Wei, Wei and Liu, Wenyu and Tian, Qi and Wang, Xinggang},
  booktitle={Proceedings of the IEEE/CVF conference on computer vision and pattern recognition},
  pages={20310--20320},
  year={2024},
  doi = {https://doi.org/10.48550/arXiv.2310.08528}}

@inproceedings{yang2024deformable,
  title={Deformable 3d gaussians for high-fidelity monocular dynamic scene reconstruction},
  author={Yang, Ziyi and Gao, Xinyu and Zhou, Wen and Jiao, Shaohui and Zhang, Yuqing and Jin, Xiaogang},
  booktitle={Proceedings of the IEEE/CVF conference on computer vision and pattern recognition},
  pages={20331--20341},
  year={2024},
  doi={https://doi.org/10.48550/arXiv.2309.13101}
}

@inproceedings{lu20243d,
  title={3d geometry-aware deformable gaussian splatting for dynamic view synthesis},
  author={Lu, Zhicheng and Guo, Xiang and Hui, Le and Chen, Tianrui and Yang, Min and Tang, Xiao and Zhu, Feng and Dai, Yuchao},
  booktitle={Proceedings of the IEEE/CVF Conference on Computer Vision and Pattern Recognition},
  pages={8900--8910},
  year={2024},
  doi = {
https://doi.org/10.48550/arXiv.2404.06270
}
}

@inproceedings{bae2024per,
  title={Per-gaussian embedding-based deformation for deformable 3d gaussian splatting},
  author={Bae, Jeongmin and Kim, Seoha and Yun, Youngsik and Lee, Hahyun and Bang, Gun and Uh, Youngjung},
  booktitle={European Conference on Computer Vision},
  pages={321--335},
  year={2024},
  organization={Springer},
  doi={
https://doi.org/10.48550/arXiv.2404.03613
Focus to learn more}
}

@inproceedings{li2024st,
  title={St-4dgs: Spatial-temporally consistent 4d gaussian splatting for efficient dynamic scene rendering},
  author={Li, Deqi and Huang, Shi-Sheng and Lu, Zhiyuan and Duan, Xinran and Huang, Hua},
  booktitle={ACM SIGGRAPH 2024 Conference Papers},
  pages={1--11},
  year={2024},
  doi={https://doi.org/10.1145/3641519.3657520}
}

@article{yang2023real,
  title={Real-time photorealistic dynamic scene representation and rendering with 4d gaussian splatting},
  author={Yang, Zeyu and Yang, Hongye and Pan, Zijie and Zhang, Li},
  journal={arXiv preprint arXiv:2310.10642},
  year={2023},
  doi={https://doi.org/10.48550/arXiv.2310.10642}
}

\appendix 
\crefalias{section}{appendix} 

\end{document}